\documentclass[twocolumn,prd,showpacs,preprintnumbers,amsmath,amssymb,aps,float,floatfix]{revtex4-1}
\usepackage{graphicx}
\usepackage{dcolumn}
\usepackage{bm}
\usepackage{multirow}
\usepackage{color}

\begin{document}
\preprint{}
\title{Shear viscosity, cavitation and hydrodynamics at LHC}
\author{Jitesh R. Bhatt{\footnote {email: jeet@prl.res.in}},
Hiranmaya Mishra{\footnote {email: hm@prl.res.in}} and
V. Sreekanth{\footnote {email: skv@prl.res.in}}}
\affiliation {Theoretical Physics Division, Physical Research Laboratory, 
Navrangpura, Ahmedabad - 380 009, India}
\date{\today}
\def\be{\begin{equation}}
\def\ee{\end{equation}}
\def\bearr{\begin{eqnarray}}
\def\eearr{\end{eqnarray}}
\def\zbf#1{{\bf {#1}}}
\def\bfm#1{\mbox{\boldmath $#1$}}
\def\hf{\frac{1}{2}}

\begin{abstract}

We study evolution of quark-gluon matter in the ultrarelativistic heavy-ion collisions
within the frame work of relativistic second-order viscous hydrodynamics.   
In particular, by using the various prescriptions of a temperature-dependent
shear viscosity to the entropy ratio, we show that 
the hydrodynamic description of the relativistic fluid becomes
invalid due to the phenomenon of cavitation. For most of the initial
conditions relevant for LHC, the cavitation sets in very early stages. 
The cavitation in this case is  entirely driven by the large values of shear viscosity. 
Moreover we also demonstrate that the conformal terms used in equations of the 
relativistic dissipative hydrodynamic can influence the cavitation time.

\end{abstract}
\pacs{25.75.-q, 25.75.Ld, 12.38.Mh, 24.10.Nz}
 \maketitle

Presently the viscosity of the strongly-interacting matter produced
in the heavy-ion collision experiments at LHC and RHIC is under
extensive investigations. The measurements of the elliptical
flow parameter $v_2$ show a strong collectivity in the fluid-flow implying 
the existence of a very low viscous stress due to shear viscosity \cite{experiments}.
According to the AdS/CFT conjecture, ratio of the shear viscosity 
to entropy density $\eta/s$ may not be lower than $1/4\pi$ which
is now known as KSS bound \cite{Kovtun:2004de}. 
It has been argued that in order to explain the 
collective flow data $\eta/s$ cannot be larger than twice
the KSS-bound \cite{Romatschke:2007mq}.

 It must be noted that the applications of the viscous hydrodynamics discussed 
above regard $\eta/s$ as independent of temperature. 
{However, recently it has been 
argued that constant $\eta/s$ is in  sharp contrast with the observed fluid behavior 
in nature where it can depend on temperature \cite{Csernai:2006zz,Niemi:2011ix}}. 
{It has been demonstrated that the temperature-dependence of $\eta/s$ can 
strongly influence the transverse momentum spectra and elliptical flow in the 
heavy-ion collision experiments at LHC \cite{Niemi:2011ix,Shen:2011eg}. 
It should be emphasized here that the ratio of bulk viscosity 
to entropy density $\zeta/s$ as a function of temperature was already 
considered by several authors 
and interesting consequences like \textit{cavitation} were studied 
\cite{Torrieri:2007fb,Rajagopal:2009yw,Bhatt:2010cy,Andreev:2011iq}. 
A similar analysis with a temperature-dependent $\eta/s$ has not been 
performed so far, which we intend to address here.
Cavitation has also been studied recently with a holographic formulation of sQGP \cite{Klimek:2011by}. 

 It is generally expected that $\eta/s$ for QGP has a minimum at the critical temperature
$T_c$, while it increases with the temperature beyond $T_c$ 
\cite{Nakamura:2004sy,Mattiello:2009db,Niemi:2011ix}. 
In this work we use $\eta/s$  prescriptions arising 
from lattice QCD (lQCD) as in Ref. \cite{Niemi:2011ix},
virial theorem type of arguments \cite{Mattiello:2009db} as well as the analytical
expressions for $\eta/s$ as given in Ref.\cite{Shen:2011eg}. 
We show that
the large values of $\eta/s$, relevant for LHC energies, can make the effective 
pressure of the fluid
very small in a time less than $2~fm/c$. This would cause 
cavitation in the fluid which in turn would limit the applicability of hydrodynamics.
It must be noted that the cavitation
at RHIC energies studied in Refs. \cite{Rajagopal:2009yw,Bhatt:2010cy,Bhatt:2011kx} 
earlier was driven by the high values of the bulk viscosity near the critical 
temperature.  However, the bulk viscosity can play an insignificant role in
 the temperatures $T>>T_c$.  In the present study we demonstrate that for 
LHC energies cavitation is solely driven by the shear viscosity. 

\begin{figure}
\includegraphics[width=8.6cm]{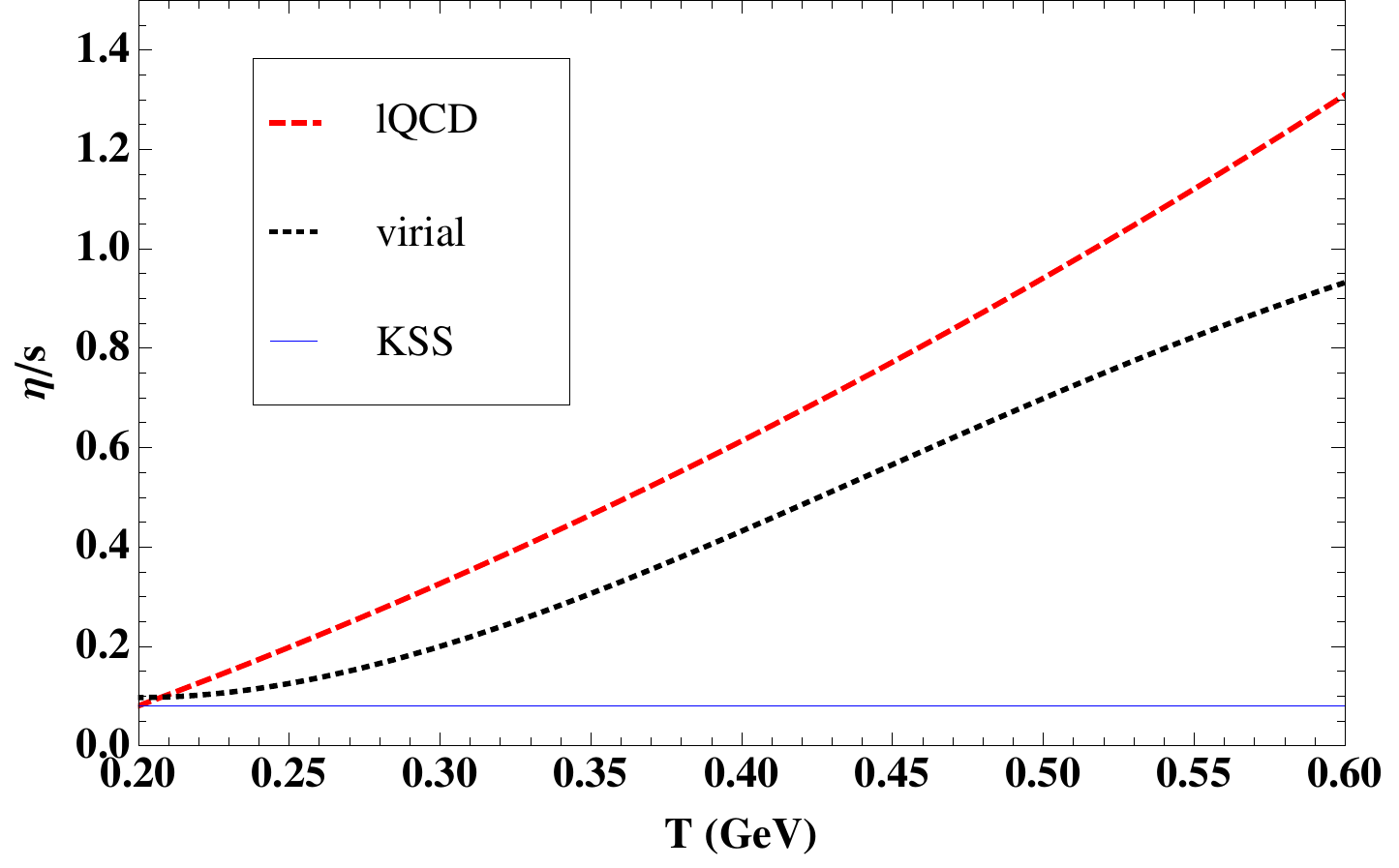}
\caption{Different prescriptions of $\eta/s$ as function of temperature, with $T_c$=0.2 GeV. 
The horizontal curve show $\eta/s=1/4\pi$ obtained from the AdS/CFT correspondence.}\label{fig:1}
\end{figure}  
 We use relativistic boost invariant causal viscous hydrodynamics equations 
in 1+1 dimensions \cite{Israel:1979wp,Bjorken:1982qr}.
One may argue against the validity of applying (1+1)-dimensional flow
in studying the relativistic heavy-ion collisions by ignoring the transverse flow. As will
be shown later for a central collision at LHC energies 
the cavitation sets during the initial stage
of the evolution in a time less than 2~fm/c. Since the transverse 
flow is negligible during the earlier stages
of a heavy-ion collision, it will not have a significant
effect on the cavitation time.
We use the parametrization of the coordinates $t=\tau$ cosh$\,\eta_s$ and $z=\tau$ sinh$\,\eta_s$, 
with the proper time $\tau = \sqrt{t^2-z^2}$ and space-time 
rapidity $\eta_s=\frac{1}{2}\,ln[\frac{t+z}{t-z}]$. Now the 4-velocity can be written as 
$u^\mu=(\rm{cosh}\,\eta_s,0,0,\rm{sinh}\,\eta_s)$.
Within the second order theory (for more details on this theory and its 
application to relativistic heavy ion collisions 
we refer to Refs. \cite{Muronga:2004sf,Romatschke:2009im,Bhatt:2010cy})
the equations dictating the longitudinal expansion of the medium are given by 
\cite{Muronga:2001zk,Heinz:2005zi,Baier:2006um,Muronga:2006zw}: 
\begin{eqnarray}
   \frac{\partial\varepsilon}{\partial\tau} &=& - \frac{1}{\tau}(\varepsilon 
  + P +\Pi - \Phi) \, ,
  \label{evo1} \\
  \frac{\partial\Phi}{\partial\tau} &=& - \frac{\Phi}{\tau_{\pi}}+\frac{2}{3}\frac
{1}{\beta_2\tau}
  -\frac{1}{\tau_{\pi}}\left[ \frac{4\tau_{\pi}}{3\tau}\Phi +\frac{\lambda_1}{2\eta^2}\Phi^2\right] 
  \, , 
  \label{evo2} \\
   \frac{\partial\Pi}{\partial\tau} &=& - \frac{\Pi}{\tau_{\Pi}} - \frac{1}{\beta_0\tau} .
\label{evo3}
\end{eqnarray}
The effects due to shear and bulk viscosity are represented via $\Phi$ and 
$\Pi$ respectively and they can contribute to the effective pressure of the fluid. 
Eqns. (\ref{evo2}-\ref{evo3}) are 
evolution equations for $\Phi$ and $\Pi$ governed by 
their relaxation times $\tau_{\pi}$ and $\tau_{\Pi}$ respectively.
Last term on the 
right-hand side of Eq. (\ref{evo2}) is due the conformal symmetry \cite{Baier:2007ix}. 
In order to close the system of Eqns. (\ref{evo1}-\ref{evo3}), one needs to use
equation of state (EoS). We have used recent lQCD results \cite{Bazavov:2009zn} for this purpose. 
At LHC energies 
the bulk viscosity 
{is expected to be negligible as $\varepsilon\approx 3P$ and one can ignore Eq. (\ref{evo3}).} 

In the local rest frame the shear stress describes the deviation from the isotropy
of the  stress tensor. To quantify this anisotropy one can define the 
longitudinal pressure $P_z$ \cite{Rajagopal:2009yw,Fries:2008ts} in absence of 
the bulk stress as 
\begin{equation}
 P_z=P-\Phi
\end{equation}
where $P$ is the equilibrium hydrodynamic pressure.

We use recent lQCD estimate for $\eta/s$ in QGP sector 
calculated by Nakamura \textit{et.~al} 
\cite{Nakamura:2004sy}. 
The  resulting $\eta/s$ from lQCD  has the expected  minimum near the critical temperature $T_c$.
It should be noted that recent lattice studies indicate a crossover rather than a 
phase-transition \cite{Aoki:2006we}. However,
for the present work this may not be an issue since we are interested in temperature 
dependence of $\eta/s$ where $T_c$ is a parameter. 
We use the parametrization of $\eta/s$ given in Ref. \cite{Denicol:2010tr}, 
where the minimum value of $\eta/s$ is $1/{4\pi}$. 
Another prescription for shear viscosity that we use is 
from Ref. \cite{Mattiello:2009db}, where 
using virial expansion techniques, the authors calculate $\eta/s$ in QGP.
Fig.[\ref{fig:1}] shows the plots of various $\eta/s$ prescriptions versus temperature with $T_c=$0.2 GeV.
The top curve shows values of $\eta/s$ obtained from the lattice results,
while the middle curve corresponds to $\eta/s$ values obtained from
the virial expansion. The horizontal line corresponds to the KSS value. 
Finally, we consider the temperature-dependent forms of $\eta/s$ 
as given in  Ref. \cite{Shen:2011eg}:
 $\left(\eta/s\right)_1 = 0.2 + 0.3\,\frac{T-T_{chem}}{T_{chem}},\, 
\left(\eta/s\right)_2 = 0.2 + 0.4\,\frac{(T-T_{chem})^2}{T^2_{chem}}\,\rm{and}\,
\left(\eta/s\right)_3 = 0.2 + 0.3\,\sqrt{\frac{T-T_{chem}}{T_{chem}}}$, 
with $T_{chem}=0.165$ GeV. 

Relaxation time $\tau_\pi=2\eta \beta_2$ can be determined
by an underlying theory other than the hydrodynamics. 
It ought to be mentioned that in the relativistic viscous hydrodynamic literature
there is some ambiguity regarding the value of the relaxation times associated 
with shear and bulk viscosity. In this work we have taken 
the relaxation time for shear viscosity 
$\tau_\pi=\frac{5\eta/s}{T}$, which is motivated by 
kinetic theory \cite{Denicol:2010tr,Niemi:2011ix}. 
In addition we also solve Eqns. (\ref{evo1}-\ref{evo2})
by taking $\tau_\pi=\frac{2\eta/s}{T}(2-\ln 2)\approx \frac{2.6\eta/s}{T}$ with $\lambda_1=\frac{\eta}{2\pi T}$, 
{inspired by results from $\mathcal N$=4 supersymmetric Yang-Mills theory 
\cite{Natsuume:2007ty,Baier:2007ix}.}

\begin{table}
\begin{center}

\begin{tabular}{cc|c|c|c|c|c|c|l}
\cline{1-8}
\multicolumn{2}{|c|}{LHC} & \multicolumn{3}{|c|}{IS ($\tau_\pi=\frac{5\eta/s}{T}$)}
 & \multicolumn{3}{|c|}{IS+C ($\tau_\pi=\frac{2.6\eta/s}{T}$)} \\ \cline{3-8} 
\multicolumn{2}{|c|}{ } & $\tau_f$& $\tau_{cav}$& $T_{cav}$
& $\tau_f$& $\tau_{cav}$& $T_{cav}$  \\ \cline{1-8} \cline{1-8}
\multicolumn{1}{|c|}{\multirow{1}{*}{$\tau_0$=0.3 fm/c}} &
\multicolumn{1}{|c|}{\scriptsize{$\eta/s$ lQCD}} & 21.18 & 0.57 & 0.421 & 14.21 &  0.52 & 0.434 &   \\ \cline{2-8}
\multicolumn{1}{|c|}{\scriptsize{$T_0$}=0.506 \scriptsize{GeV}}                        &
\multicolumn{1}{|c|}{\scriptsize{$\eta/s$ virial}} & 20.61 & 0.63 & 0.410 & 13.93 &   0.93 & 0.374 &   \\ \cline{1-8} \cline{1-8}

\multicolumn{1}{|c|}{\multirow{1}{*}{$\tau_0$=0.3 fm/c}} &
\multicolumn{1}{|c|}{\scriptsize{$\eta/s$ lQCD}} & 31.58 & 0.57 & 0.465 & 20.31 &  0.52 & 0.479 &   \\ \cline{2-8}
\multicolumn{1}{|c|}{\scriptsize{$T_0$}=0.560 \scriptsize{GeV}}                        &
\multicolumn{1}{|c|}{\scriptsize{$\eta/s$ virial}} & 25.06 & 0.68 & 0.444 & 18.12 &   1.20 & 0.385 &   \\ \cline{1-8}  \cline{1-8}

\multicolumn{1}{|c|}{\multirow{1}{*}{$\tau_0$=0.6 fm/c}} &
\multicolumn{1}{|c|}{\scriptsize{$\eta/s$ lQCD}} & 12.40 & 1.20 & 0.333 & 10.83 &  1.53 & 0.316 &   \\ \cline{2-8}
\multicolumn{1}{|c|}{\scriptsize{$T_0$}=0.405 \scriptsize{GeV}}                        &
\multicolumn{1}{|c|}{\scriptsize{$\eta/s$ virial}} & 15.30 & 1.27 & 0.329 & 11.88 &   - & - &   \\ \cline{1-8}  \cline{1-8}

\multicolumn{1}{|c|}{\multirow{1}{*}{$\tau_0$=0.6 fm/c}} &
\multicolumn{1}{|c|}{\scriptsize{$\eta/s$ lQCD}} & 18.36 & 1.21 & 0.369 & 15.63 & 1.29 & 0.365 &   \\ \cline{2-8}
\multicolumn{1}{|c|}{\scriptsize{$T_0$}=0.450 \scriptsize{GeV}}                        &
\multicolumn{1}{|c|}{\scriptsize{$\eta/s$ virial}} & 19.84 & 1.43 & 0.353 & 16.07 &   - & - &   \\ \cline{1-8}  \cline{1-8}

\multicolumn{1}{|c|}{\multirow{1}{*}{$\tau_0$=1.0 fm/c}} &
\multicolumn{1}{|c|}{\scriptsize{$\eta/s$ lQCD}} & 10.48 & - & - & 9.98 & - & - &   \\ \cline{2-8}
\multicolumn{1}{|c|}{\scriptsize{$T_0$}=0.350 \scriptsize{GeV}}                        &
\multicolumn{1}{|c|}{\scriptsize{$\eta/s$ virial}} & 13.04 & 2.16 & 0.283 & 11.17 &   - & - &   \\ \cline{1-8}  \cline{1-8} 
\end{tabular}
 
\end{center}
\caption{
Column IS corresponds to the case when the conformal terms are neglected from
the hydrodynamics equations. In this case the relaxation time $\tau_\pi$ 
from the kinetic theory is taken in to account. The column IS+C corresponds to the case when the
conformal terms and  $\tau_\pi$ obtained from the supersymmetric Yang-Mills theory
are included in the equations of hydrodynamics. The cavitation time $\tau_{cav}$ and
$\tau_f$ are measured in the unit of fm/c and the cavitation temperature $T_{cav}$
is shown in the units of GeV. $\tau_{cav}$ and $T_{cav}$ are left blank when there is no cavitation.}
\end{table}

Next we present the numerical solutions for the equations of hydrodynamics.
First we consider the case with temperature-dependent $\eta/s$ taken from lQCD calculations. 
Fig.[\ref{fig-clQCD}] shows the plots of longitudinal pressure $P_z$ versus the proper time for the cases of 
pure Israel-Stewart type (IS) hydro by neglecting the conformal terms in Eq. (\ref{evo2}) and 
with conformal terms (IS+C). In the case of IS we use $\tau_\pi$ from the kinetic theory and 
from the supersymmetric Yang-Mills theory when we consider IS+C case. 
We plot $P_z$ for these two cases with the initial temperatures 0.405 and 0.450 GeV. 
The starting time $\tau_0$ is chosen to be 0.6 fm/c. Let us first consider the case with $T_0=0.405$ GeV. 
From the figure it is clear that longitudinal pressure becomes negative in the IS case around cavitation 
time $\tau_{cav}=$~1.20~fm/c. 
The temperature $T_{cav}$ at which the cavitation occurs is about 0.333~GeV
which is much larger than the critical temperature $T_c$ .
Thus the cavitation can take place
very early during the evolution. This, we believe, provides {\it a posteriori}
justification for neglecting the transverse flow; as the hydrodynamic
treatment may not be valid for the time larger than $\tau_{cav}$. 
Further, if we include the conformal terms in Eq. (\ref{evo2}) together
with the relaxation time obtained from supersymmetric Yang-Mills (IS+C), 
the cavitation time increases marginally and becomes $\tau_{cav}$=1.53~fm/c. 
Similarly $T_{cav}$=0.316~GeV is less than the cavitation temperature
without the conformal terms. 
Next we consider a higher initial temperature $T_0=0.450$ GeV. Here also we observe cavitation 
for both IS and IS+C cases as in the previous case with $T_0=0.405$ GeV. 
For IS case cavitation happens at a time $\tau_{cav}=1.21$ fm/c which is only marginally greater than the 
corresponding $T_0=0.405$ GeV case considered previously. However, here the temperature at which cavitation occurs is higher 
($T_{cav}=0.369$~GeV) than the previous case. This difference is expected since the initial temperature for the latter case is 
also larger. IS+C case with $T_0=0.450$ GeV, cavitation sets 
in at $\tau_{cav}=1.29$ fm/c with $T_{cav}$=0.365~GeV. Again we note that 
there is not much difference between the cavitation times in IS and IS+C cases.

\begin{figure}
 \includegraphics[width=8.6cm]{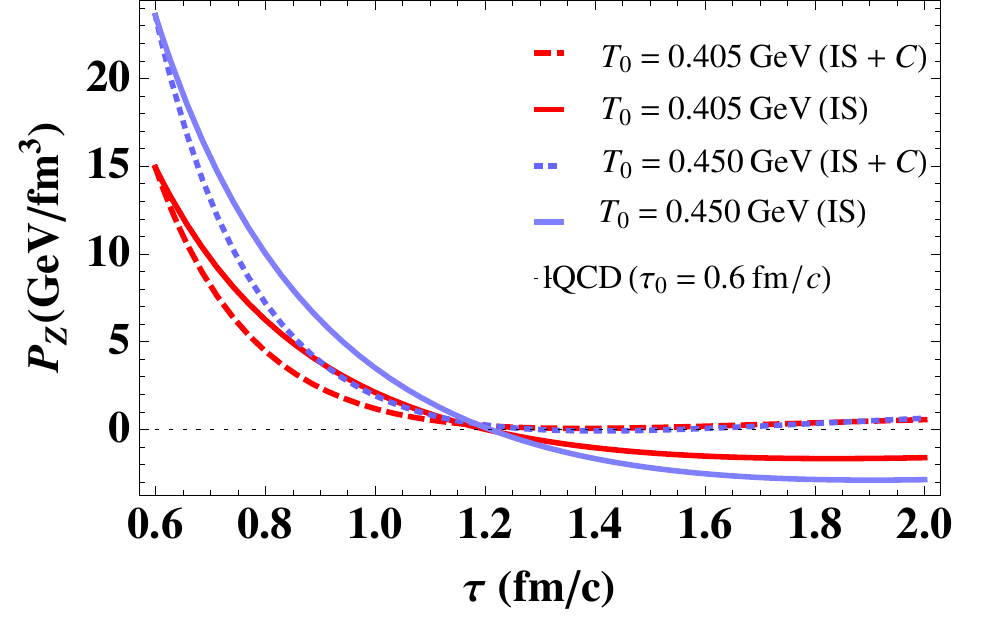}
\caption{The longitudinal pressure $P_z$ as function of time for IS and IS+C hydrodynamics. 
Initial time is taken to be 0.6 fm/c with initial temperatures 0.405 and 0.450 GeV. 
$\eta/s(T)$ is obtained from the lQCD curve shown in Fig.[\ref{fig:1}].}\label{fig-clQCD}
\end{figure}

In Fig.[\ref{fig-cvirial}], we show $P_z$ as function of time by taking $\eta/s$
values using the virial expansion techniques given in Ref. \cite{Mattiello:2009db}. Values for  $\tau_0$ and $T_0$
are same as in Fig.[\ref{fig-clQCD}]. Here in the IS case with $T_0=0.450$ GeV we can see that 
cavitation sets in around $1.43$ fm/c when the system temperature is $0.353$ GeV. However, as one can see from 
Fig.[\ref{fig-cvirial}], when we include conformal terms (IS+C case) cavitation scenario is avoided. 
Next we lower the initial temperature to $0.405$ GeV and consider the IS case. Here system reaches a negative 
longitudinal pressure stage at $\tau_{cav}=1.27$~fm/c with $T_{cav}=0.329$ GeV. But with conformal terms included, 
as one can see from the figure, the longitudinal pressure remains positive although it assumes a very small 
value by $2$ fm/c. Since the values of $\eta/s$ for the virial expansion techniques are systematically smaller 
than $\eta/s$ for the lQCD results as shown in Fig.[\ref{fig:1}], the corresponding
cavitation time is larger than that shown in Fig.[\ref{fig-clQCD}]. However, the cavitation
temperature $T_{cav}$ is smaller than the corresponding cases discussed in Fig.[\ref{fig-clQCD}].

\begin{figure}
 \includegraphics[width=8.6cm]{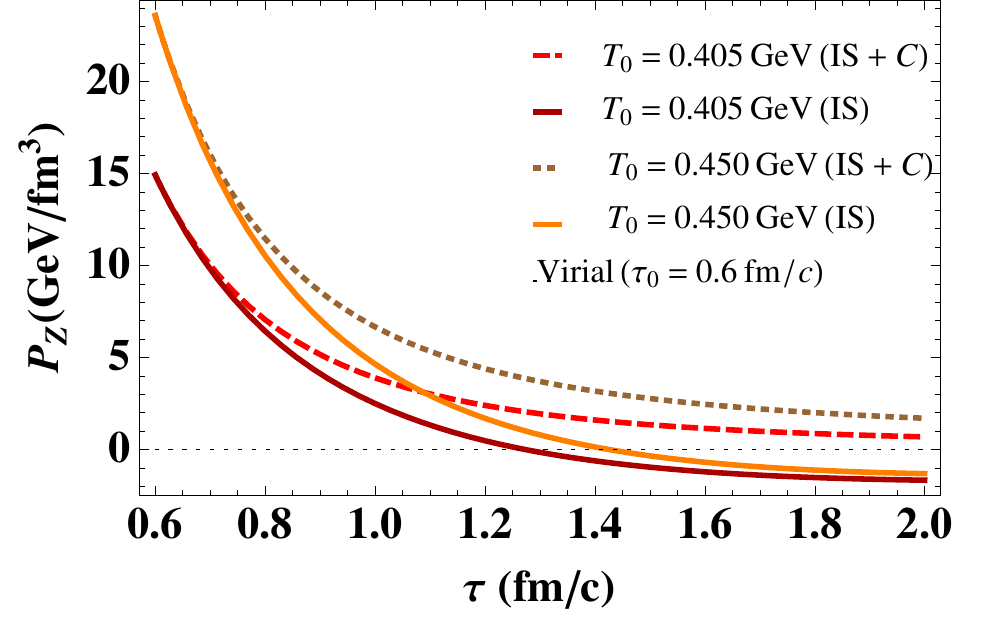}
\caption{The longitudinal pressure $P_z$ as function of time for IS and IS+C hydrodynamics. 
Initial time is taken to be 0.6 fm/c with initial temperatures 0.405 and 0.450 GeV. 
$\eta/s(T)$ taken from the virial expansion 
techniques curve in Fig.[\ref{fig:1}].}\label{fig-cvirial}
\end{figure}

Further, we have changed the values of the initial time by considering the
case $\tau_0=0.3$~fm/c and $\tau_0=1.0$~fm/c. These results are summarized in
Table I. For $\tau_0=0.3$~fm/c and $T_0=0.560$~GeV case, the cavitation occurs around $\tau_{cav}=0.6$~fm/c
for the lQCD $\eta/s$ while it occurs around $\tau_{cav}=0.68$~fm/c for
$\eta/s$ obtained from the virial expansion. 
For the case with $\tau_0=1.0$~fm/c and $T_0=0.350$~GeV, for $\eta/s$ from virial expansion,
the cavitation occurs 
 around $\tau_{cav}=2.16$~fm/c. However, in this case when the $\eta/s$ values from
lQCD are used there is no cavitation.
We would like to note that the table shows no entries for $\tau_{cav}$ and $T_{cav}$
for certain cases. For such instances
the longitudinal pressure remains positive 
and there is no cavitation. Table I indicates for the given initial conditions
there are  more number of  no-cavitation instances when the conformal terms
in the equations of the hydrodynamics are taken into account.

We also summarise the results for $\tau_f$, the total time taken by the system to reach $T_c$ by ignoring the cavitation in Table I.
One can see that with $T_0=0.405$ GeV for lQCD (virial) case $\tau_f$=12.40 (15.30) fm/c without the conformal
term and $\tau_f$=10.83 (11.88)~fm/c if the term is included. 
Thus the inclusion
of the conformal terms reduces $\tau_f$. 
We would like to emphasize that in this work we have taken
a rather conservative  initial value  $\Phi(\tau_0)$=0 so that the initial
value of the longitudinal pressure is always positive \cite{Martinez:2009mf}. 
Instead if one includes the first-order (Navier-Stokes) initial value $\Phi(\tau_0)=4\eta(T_0)/(3\tau_0)$,
then the cavitation can occur at even earlier time and higher temperature.

\begin{figure}
 \includegraphics[width=8.6cm]{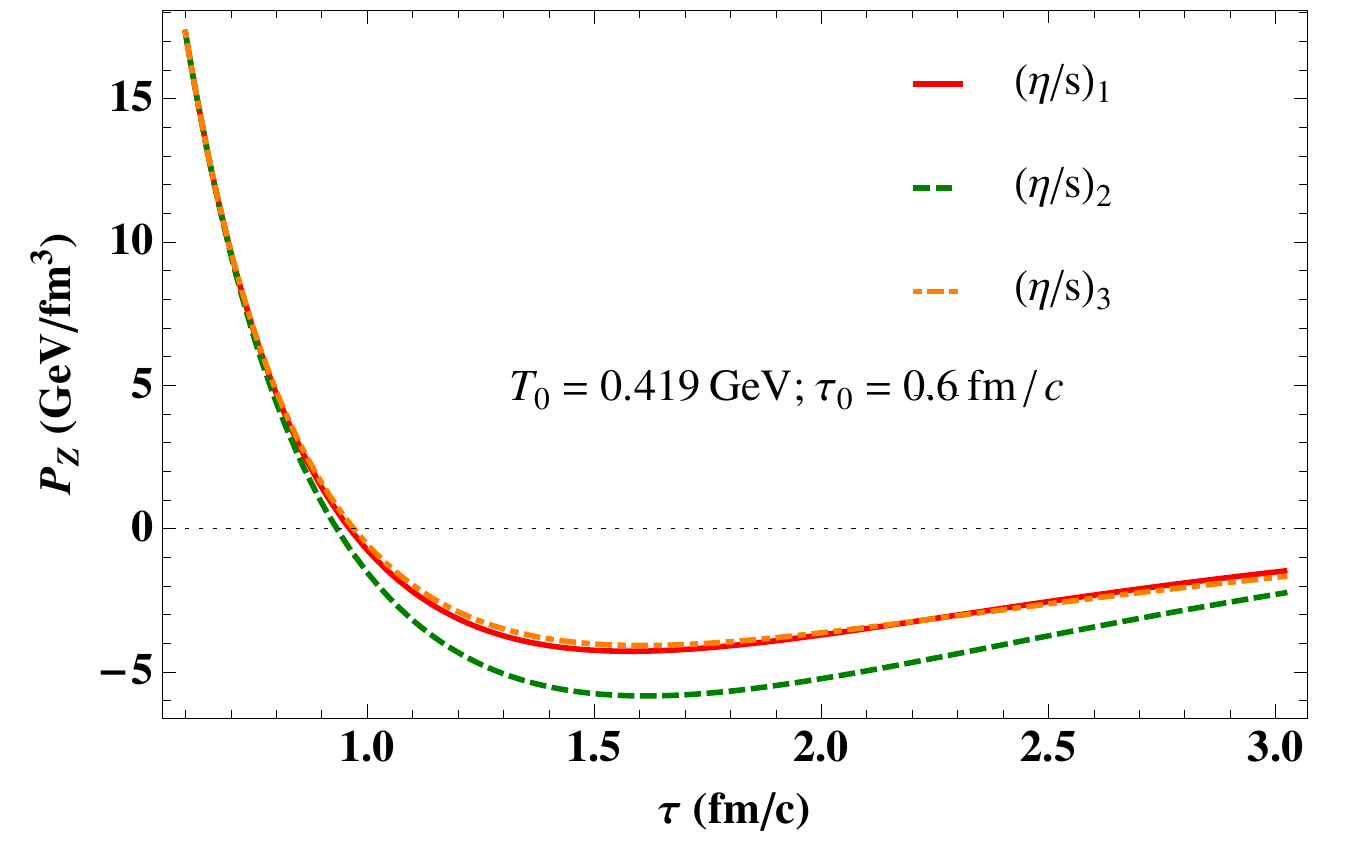}
\caption{Cavitation with various $\eta/s$ prescriptions considered by Shen \textit{et.al.} in Ref. \cite{Shen:2011eg}. 
The initial temperature is taken to be 0.419~GeV with initial time $0.6$ fm/c.}\label{Fig.5}
\end{figure}

Next, we repeat our analysis using the temperature-dependent $\eta/s$
 prescriptions given 
in Ref. \cite{Shen:2011eg}.
With the same initial conditions as in Ref. \cite{Shen:2011eg} we find
that  the longitudinal pressure becomes negative
very early $\sim 1$ fm/c for $all$ the cases they have considered. 
Fig.[\ref{Fig.5}] shows  $P_z$  versus $\tau$ for initial temperature
$T_0=0.419$ GeV and $\tau_0=0.6$ fm/c.
In this case also cavitation sets in early in about $\tau_{cav}\sim$ 1~fm/c.

Further, we also consider $\eta/s=1/4\pi$ and $\zeta/s$ as function of T
as in Ref. \cite{Rajagopal:2009yw} for LHC energies. It is found that the cavitation 
does not occur in this case unlike the results for RHIC energies 
\cite{Rajagopal:2009yw,Bhatt:2010cy,Bhatt:2011kx}.
One may naively expect that when the  system temperature
reach  $T\sim T_c$, the bulk viscosity become large enough to drive cavitation. 
However, the cavitation occurs when the viscous stress ($\Pi$ and/or $\Phi$)
 has a peak in its temporal
profile (see Fig.[\ref{Fig.ano}]). The height of the peak is determined by $\tau_0$, $T_0$ and the
initial values of $\zeta/s$ or $\eta/s$. For LHC energies, we find that even at
 the peak value of the viscous stress $\Pi$, the condition $\Pi< P$ is 
satisfied and therefore cavitation does not occur.

\begin{figure}
 \includegraphics[width=8.6cm]{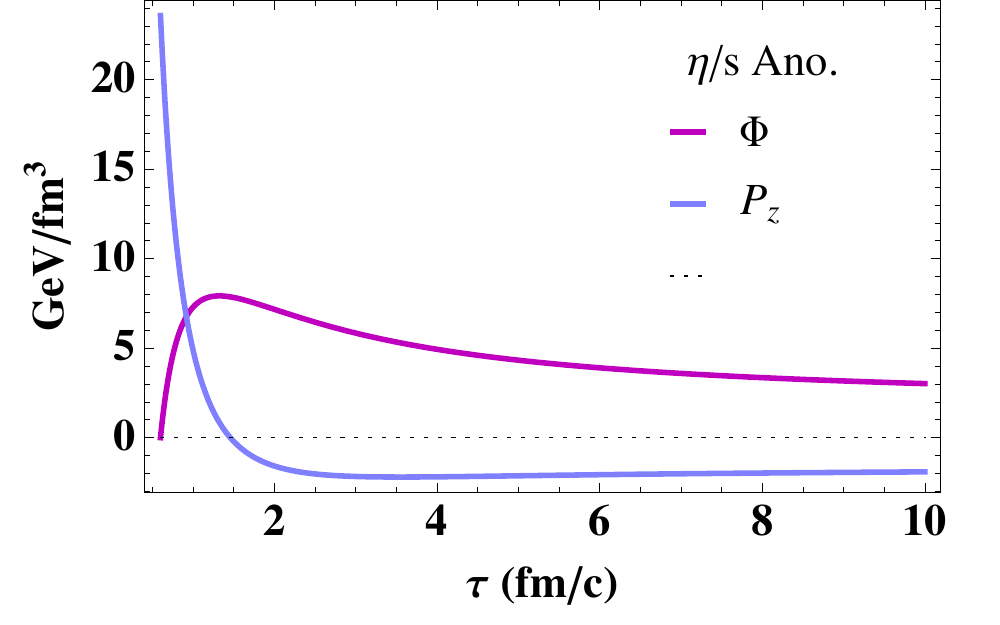}
\caption{Cavitation along with anomalous viscosity. The longitudinal pressure $P_z$ and $\Phi$ as function of
time. The initial temperature is taken to be 0.450~GeV with initial time $0.6$ fm/c.}\label{Fig.ano}
\end{figure}

We have further considered the effect of anomalous viscosity ($\eta_A$), 
which may be important during the early time evolution in the 
hydrodynamics \cite{Asakawa:2006tc}. 
We use an effective shear viscosity $\eta^{-1}=\eta_A^{-1} + \eta_C^{-1}$ as discussed in Ref. \cite{Asakawa:2006tc}. 
Here, $\eta_C$ the collisional viscosity is taken from lQCD and for $\eta_A/s$ we use the expression from Ref. \cite{Asakawa:2006tc}. 
In this case, (with $\tau_0=0.6$ fm/c and $T_0=450$ MeV),
cavitation sets in at a time $1.46$ fm/c when the system is at a temperature $351$ MeV.
The initial value of anomalous viscosity to entropy density ratio is $\sim0.23$. 
 The results are presented in Fig.[\ref{Fig.ano}], 
where we plot the shear stress term $\Phi$ and longitudinal pressure $P_z$ 
as a function of proper time. 
{As is clear from Fig.[\ref{Fig.ano}] the shear stress $\Phi$ increases sharply from its
initial value. The maximum value of $\Phi$ and the time it
takes to reach that value strongly depend upon $\tau_\pi$.} 
This sharp rise of $\Phi$ result in a sharp reduction of $P_z$, which, 
finally becomes negative at $\tau_{cav}$.
 
Thus to summarise, we have shown by using various prescriptions for
a temperature-dependent 
$\eta/s$ that at LHC energies the  higher values of shear stresses
can induce the cavitation. 
This will in turn make the hydrodynamic
treatment invalid beyond cavitation time $\tau_{cav}$. 
We have studied shear viscosity induced cavitation using one dimensional boost invariant 
causal dissipative hydrodynamics of Israel-Stewart. 
One would of course like to do an analysis using a (3+1)-dimensional 
viscous hydrodynamics like e.g. in Ref. \cite{Schenke:2010rr}. 
Since cavitation occurs during the early stages of the collision, 
we believe that the inclusion of transverse flow will not alter the result qualitatively. 
However, as a caveat, we would like to mention that
the difference between the initial conditions for  
the ``cavitation" and ``no-cavitation" cases is rather small, see Table I. It 
remains to be seen if the inclusion of transverse flow can alter the cavitation
scenario in a qualitative way. 
It is worth noting here the negative pressure scenario may be circumvented
by considering anisotropic corrections in the distribution functions 
\cite{Martinez:2010sc}. It should be emphasized that there exist 
alternate formulations of dissipative relativistic fluid dynamics 
where the longitudinal pressure remains positive e.g. in Ref. \cite{Denicol:2010xn}
It has been shown recently that the inclusion of the cavitation condition in 
boost invariant hydrodynamics can change the particle spectra from expanding QGP 
\cite{Bhatt:2010cy,Bhatt:2011kx}. 
Based on the various prescriptions of $\eta/s $ our results indicate that the
hydrodynamical description is valid about $\tau_{cav}\approx2$~fm/c at
LHC energies. 
Beyond $\tau_{cav}$, the fluid might fragment \cite{Torrieri:2007fb} or form inhomogeneous clusters.
Let us note that one of the assumptions of the statistical hadronisation models
lies in creation of  extended clusters of quark matter which hadronize
statistically \cite{Becattini:2009sc}. 
Alternately, as has been attempted recently one can possibly
use a hybrid approach for the description of fire ball expansion
applying viscous hydrodynamics for the QGP stage and
then coupling it to a microscopic kinetic evolution for the hadronic stage 
\cite{Song:2010aq}.
Mere integration of the equations of hydrodynamics may not tell us 
about cavitation. We therefore believe that the conditions 
for cavitation may be required to be incorporated in the hydrodynamical
codes.

\acknowledgements 

One of the authors (VS) would like to thank P. Romatschke for many useful suggestions.
JRB would like to thank E. Shuryak for useful discussions.
We would like to thank K. Rajagopal for a critical reading of the manuscript and suggestions. 
We would like to thank G. Denicol and T. Kodama for 
providing their parametrisation of the lQCD results for $\eta/s$.

\end{document}